\documentclass{tcibook}
\usepackage{fancyhea}
\usepackage{work}
\usepackage{bm}       
\usepackage{graphicx}
\usepackage{hyperref}      

\newcommand{\nc}{\newcommand}  

\def\beq{\begin{equation}}
\def\eeq#1{\label{#1}\end{equation}}
\def\eeqn{\end{equation}}

\newenvironment{Eqnarray}%
   {\arraycolsep 0.14em\begin{eqnarray}}{\end{eqnarray}}
\def\beqa{\begin{Eqnarray}}
\def\eeqa#1{\label{#1}\end{Eqnarray}}
\def\eeqan{\end{Eqnarray}}

\nc{\ra}{\rightarrow}  
\nc{\slsh}{\slash\hspace*{-0.22cm}}
\def\Re{{\cal R \mskip-4mu \lower.1ex \hbox{\it e}\,}}
\def\Im{{\cal I \mskip-5mu \lower.1ex \hbox{\it m}\,}}

\nc{\vev}[1]{ \left\langle {#1} \right\rangle }
\nc{\bra}[1]{ \langle {#1} | }
\nc{\ket}[1]{ | {#1} \rangle }
\nc{\fb}{\,{\rm fb}^{-1}}
\nc{\ev}{{\rm eV}}
\nc{\kev}{{\rm keV}}
\nc{\Mev}{{\rm MeV}}
\nc{\gev}{{\rm GeV}}
\nc{\tev}{{\rm TeV}}
\nc{\mev}{{\rm MeV}}

\def\del{\partial}
\def\Dslash{\not{\hbox{\kern-4pt $D$}}}
\def\dslash{\not{\hbox{\kern-2pt $\del$}}}
\def\pslash{\not{\hbox{\kern-2pt $p$}}}
\def\ETmiss{ \not{\hbox{\kern-4pt $E$}}_T }

\def\msb{{\bar{\ssstyle M \kern -1pt S}}}

\begin{document}

\def\bibname{References}
\bibliographystyle{plain}

\raggedbottom

\pagenumbering{roman}

\parindent=0pt
\parskip=8pt
\setlength{\evensidemargin}{0pt}
\setlength{\oddsidemargin}{0pt}
\setlength{\marginparsep}{0.0in}
\setlength{\marginparwidth}{0.0in}
\marginparpush=0pt


\pagenumbering{arabic}

\newcommand {\met}{\ensuremath{E^{\rm miss}_{\rm T}}}
\newcommand {\pt}{\ensuremath{p_{\rm T}}}
\newcommand {\hadt}{\ensuremath{H_{\rm T}}}
\newcommand {\hgt}{\ensuremath{H^{*}_{\rm T}}}

\renewcommand{\chapname}{chap:intro_}
\renewcommand{\chapterdir}{.}
\renewcommand{\arraystretch}{1.25}
\addtolength{\arraycolsep}{-3pt}


\begin{boldmath}
\chapter{Excited quark production at a 100\,TeV VLHC}
\label{chap:VLHC_Qstar}
\end{boldmath}

\begin{center}

{Jacob Anderson}

Fermi National Accelerator Laboratory, Batavai, IL USA\\

\end{center}

I look for a dijet resonance produced by an excited quark $q^*$ in a
simulated sample corresponding to 3\,ab$^{-1}$ of $pp$ collisions at
$\sqrt{s} = 100$\,TeV.  Using a cut and count analysis approach, I
demonstrate the potential to explore $q^*$ masses up to 50\,TeV,
corresponding to a length scale of around 4\,am.

\section{Introduction}

At the new energy regime afforded by a $\sqrt{s} =100$\,TeV $pp$
collider (VLHC)~\cite{Ambrosio:2001ej}, it should be possible to
search for structure of the quarks of the standard model
\cite{Baur:1989kv, Baur:1987ga, Harris:1996ct}.  The primary
background for a signal of this type is the QCD dijet production.

\section{Data and detector simulation}

Using \textsc{pythia}~\cite{Sjostrand:2006za} and the CTEQ6L1 PDF's, I
genererate both background and signal events and perform the parton
showering.  The center of mass energy of the $pp$ collisions is
100\,TeV.  The ``Snowmass'' detector~\cite{Anderson:2013kda}, based on
the Delphes~\cite{Delphes} detector simulation package, provides the
detector response.  The foreseen rate for ``pile-up'' $pp$
interactions is an average of 140 minimum bias interactions per bunch
crossing.  The integrated liminosity is expected to reach
3\,ab$^{-1}$.  During the production of the signal samples, I use the
convention $\Lambda = m_{q^*}$.  The cross-sections reported by
\textsc{pythia} for the considered processes are contained in
Table~\ref{tab:qstar_xsec}.  As in~\cite{Baur:1989kv} the branching
fraction to dijets is 85\%.

\begin{table}
\caption{\label{tab:qstar_xsec} Cross-sections for signal processes.}
\begin{center}
\begin{tabular}{|c|c|}
\hline
$q^*$ mass & cross-sections (fb) \\
\hline
30\,TeV & 22.8 \\
40\,TeV & 0.986 \\
50\,TeV & $4.38\times 10^{-2}$ \\
60\,TeV & $2.22\times 10^{-3}$ \\
\hline
\end{tabular}
\end{center}
\end{table}

\section{Analysis}

From the signal and background samples, I select the two jets with the
largest $p_\mathrm{T}$.  To reduce the relative contribution of the
QCD dijet production, I require that the jet $p_\mathrm{T} > 10$\,TeV
and $|\eta | < 1.0$.  I use the dijet invariant mass spectrum to
separate the signal and background components.  The mass spectra for
the background and signals with masses from 30\,TeV to 60\,TeV are
shown in Fig.~\ref{fig:qstar_vlhc}(left).

Because of the large cross-section for the QCD process, I parameterize
the spectral shape from a large sample of simulated events and scale
this shape to provide the background estimate.  The functional form
selected for the background is an exponential decay with an error
function turn-on at the low-mass end of the specturm.  The line-shapes
for the various signal hypotheses are parameterized by a Gaussian
distribution for the peak plus a wide structure to capture poorly
reconstructed events.

Limits are derived using a simple counting method.  I select a signal
window corresponding to $\pm 2\sigma$ of the peaking component of the
line-shape.  Using the signal and background counts within the signal
window, I place an asymptotic 95\% CL upper limit on the cross-section
for a given $q^*$ mass.  Systematic uncertainties are not included in
the limits.  Given the simplicity of the analysis, degradations due to
systematic uncertainties can likely be compensated through more
advanced analysis techniques.

\section{Results}

\begin{figure}
\begin{center}
\includegraphics[width=0.475\textwidth]{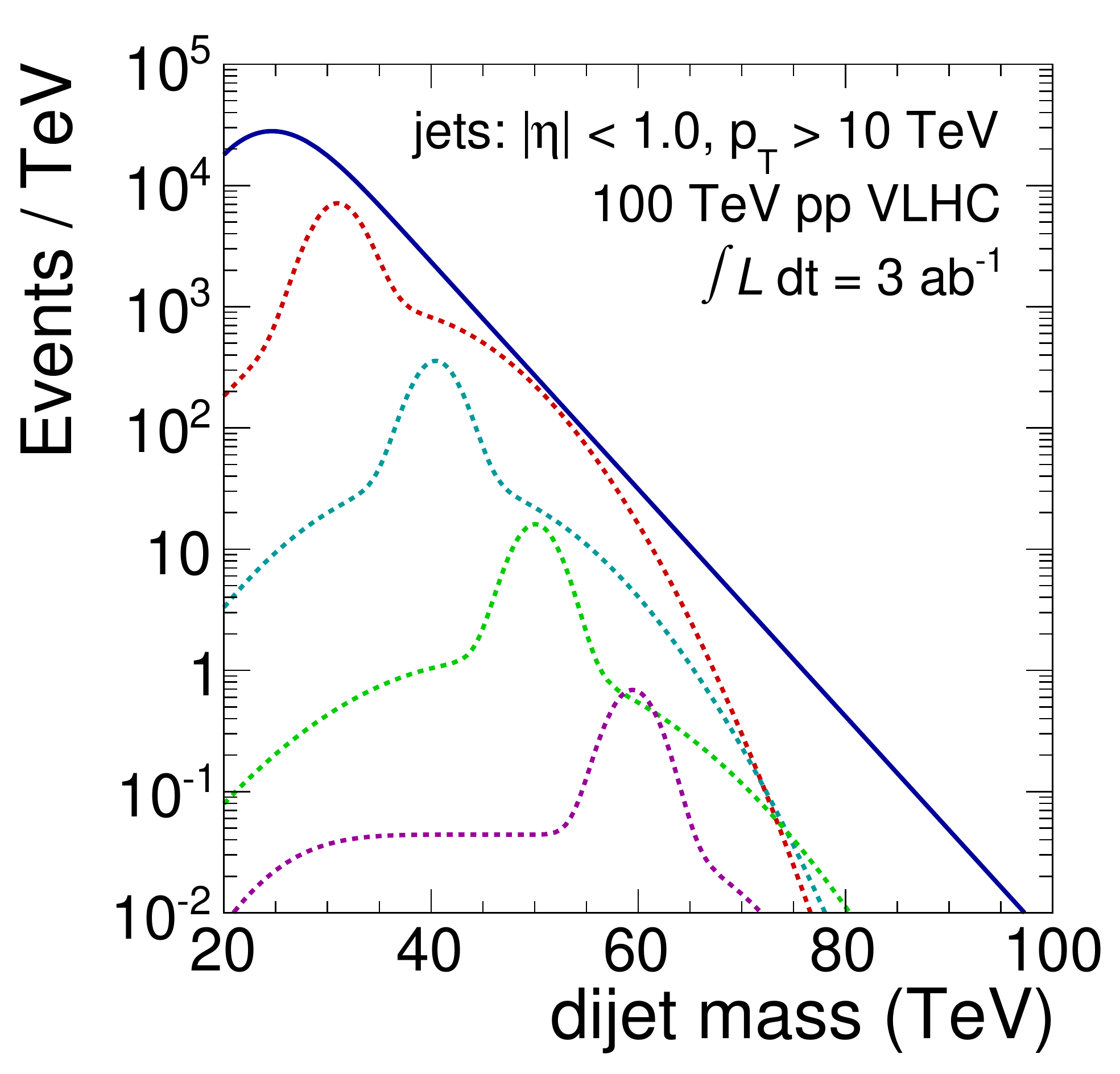}
\includegraphics[width=0.475\textwidth]{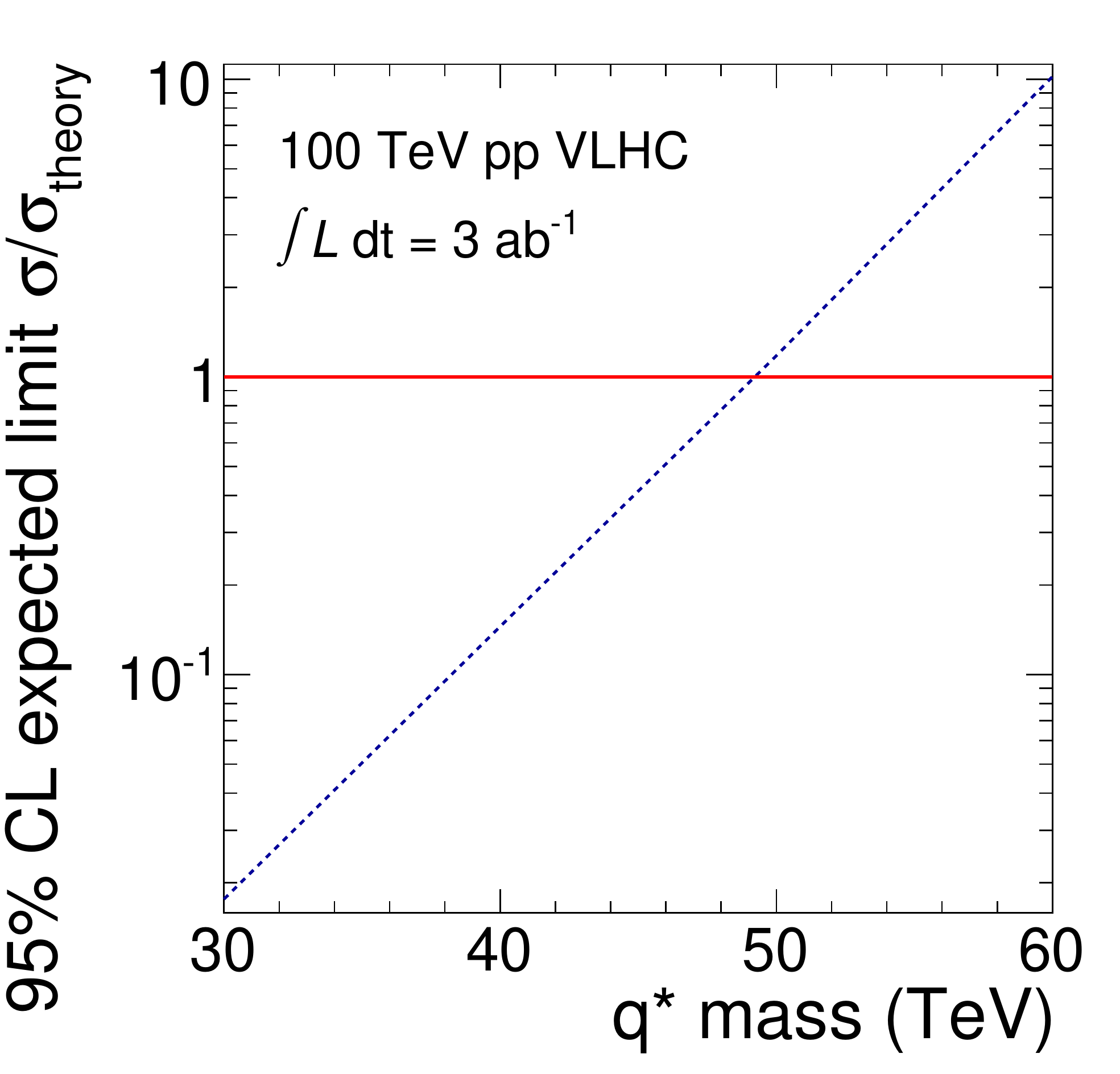}
\end{center}
\caption{\label{fig:qstar_vlhc}(left) Dijet mass spectrum for
  background dijet production (solid blue line) and $q^*$ models with
  masses from 30\,TeV to 60\,TeV (dashed colored lines). (right) 95\%
  CL upperlimit on the cross-section of $q^*$ resonance as a function
  of its mass normalized by the theoretical prediction for the
  cross-section.}
\end{figure}

Figure~\ref{fig:qstar_vlhc}(right) shows the expected limit as a
function of $q^*$ mass.  The full 3\,ab$^{-1}$ of integrated
luminosity should allow the exploration of $q^*$ masses up to 50\,TeV
at the VLHC.  A reach of 50\,TeV would study substructure within
quarks at the scale of 4\,am.



\end{document}